\input{aipcheck}

\documentclass[
    ,final            
  ]
  {aipproc}

\layoutstyle{6x9}

\newcommand{\apj}{ApJ}
\newcommand{\apjl}{ApJ Letters}

\newcommand\etal{\textit{et al.\ }}

\def\MSUN{\rm M_{\odot}}

\def\MSUNS{\rm M_{\odot}\,s^{-1}}

\newbox\grsign \setbox\grsign=\hbox{$>$} \newdimen\grdimen \grdimen=\ht\grsign
\newbox\simlessbox \newbox\simgreatbox
\setbox\simgreatbox=\hbox{\raise.5ex\hbox{$>$}\llap
     {\lower.5ex\hbox{$\sim$}}}\ht1=\grdimen\dp1=0pt
\setbox\simlessbox=\hbox{\raise.5ex\hbox{$<$}\llap
     {\lower.5ex\hbox{$\sim$}}}\ht2=\grdimen\dp2=0pt
\def\simgreat{\mathrel{\copy\simgreatbox}}

\begin{document}

\title{MHD simulations of the collapsar model for GRBs.}

\author{Daniel Proga}{
  address={JILA, University of Colorado, Boulder, CO 80309-0440, USA}
}

\author{Andrew I. MacFadyen}{
  address={California Institute of Technology, Mail Code 130-33, 
Pasadena, CA 9112}
}

\author{Philip J. Armitage}{
  address={JILA, University of Colorado, Boulder, CO 80309-0440, USA}
}

\author{Mitchell C. Begelman}{
  address={JILA, University of Colorado, Boulder, CO 80309-0440, USA}
}

\begin{abstract}
We present results from axisymmetric, time-dependent
magnetohydrodynamic (MHD) simulations of the collapsar model for
gamma-ray bursts.  Our main conclusion is that, within
the collapsar model, MHD effects alone are able to launch, accelerate
and sustain a strong polar outflow.  We also find that the outflow is
Poynting flux-dominated, and note that this provides favorable initial
conditions for the subsequent production of a baryon-poor fireball.
\end{abstract}

\maketitle


\section{Introduction}

The collapsar model is one of most promising scenarios to explain the
huge release of energy in a matter of seconds, associated with
gamma-ray bursts (GRBs; \cite{Woosley}; \cite{P98}; \cite{MW}; 
\cite{Pop}; \cite{MWH}). In this model, the collapsed iron
core of a massive star accretes gas at a high rate ($\sim 1 \MSUNS$)
producing a large neutrino flux, a powerful outflow, and a GRB.
Although the association of long duration GRBs with stellar
collapse is now secure (\cite{hjorth03}; \cite{Staneketal}),
basic properties of the central engine are uncertain. In part, this is
because previous numerical studies of the collapsar model did not
explicitly include magnetic fields, although they are commonly
accepted as a key element of accretion flows and outflows.

We present a study of the time evolution of 2.5-dimensional,
magnetohydrodynamic (MHD) flows in the collapsar model. This study is
an extension of existing models of MHD accretion flows onto a black
hole (BH; \cite{PB03}).  In particular, we
include a realistic equation of state, photodisintegration of
bound nuclei and cooling due to neutrino emission.  Our study is also
an extension of collapsar simulations by \cite{MW}, as we consider very similar
neutrino physics and initial conditions but solve MHD instead of
hydrodynamical equations.

\section{Models}

We begin the simulations after the $1.7~\MSUN$ iron
core of a 25~$\MSUN$ presupernova star has collapsed and study the
ensuing accretion of the $7~\MSUN$ helium envelope onto the central
black hole formed by the collapsed iron core.  We consider a
spherically symmetric progenitor model, but with spherical symmetry
broken by the introduction of a small, latitude-dependent angular
momentum and a weak radial magnetic field.
For more details, see \cite{Petal}). 

\section{Results}

We find that after a transient episode of infall, lasting 0.13 s, the
gas with $l\simgreat 2 R_S c$ piles up outside the black hole and
forms a thick torus bounded by a centrifugal barrier near the rotation
axis.  Soon after the torus forms (i.e., within a couple of orbital
 times at
the inner edge), the magnetic field is amplified by the magnetorotational
instability (MRI, e.g., \cite{BH91}) and shear.
We have verified that most of the inner torus is unstable to MRI, and
that our simulations have enough resolution to resolve, albeit
marginally, the fastest growing MRI mode.  The torus starts evolving
rapidly and accretes onto the black hole. Another important effect of
magnetic fields is that the torus produces a magnetized corona and an
outflow.  The presence of the corona and outflow is essential to the
evolution of the inner flow at all times and the entire flow close to
the rotational axis during the latter phase of the evolution.  We find
that the outflow very quickly becomes sufficiently strong to overcome
supersonically infalling gas (the radial Mach number in the polar
funnel near the inner radius is $\sim 5$) and makes its way outward,
reaching the outer boundary at $t=0.25$ s. Due to limited computing
time, our simulations were stopped at $t=0.28215$~s, which corresponds
to 6705 orbits of the flow near the inner boundary.  We
expect the accretion to continue much longer, roughly the collapse
timescale of the Helium core ($\sim 10$~s), as in \cite{MW}.

\begin{figure}
  \includegraphics[bb=-50 270 500 100, width=.6\textwidth, angle=90]{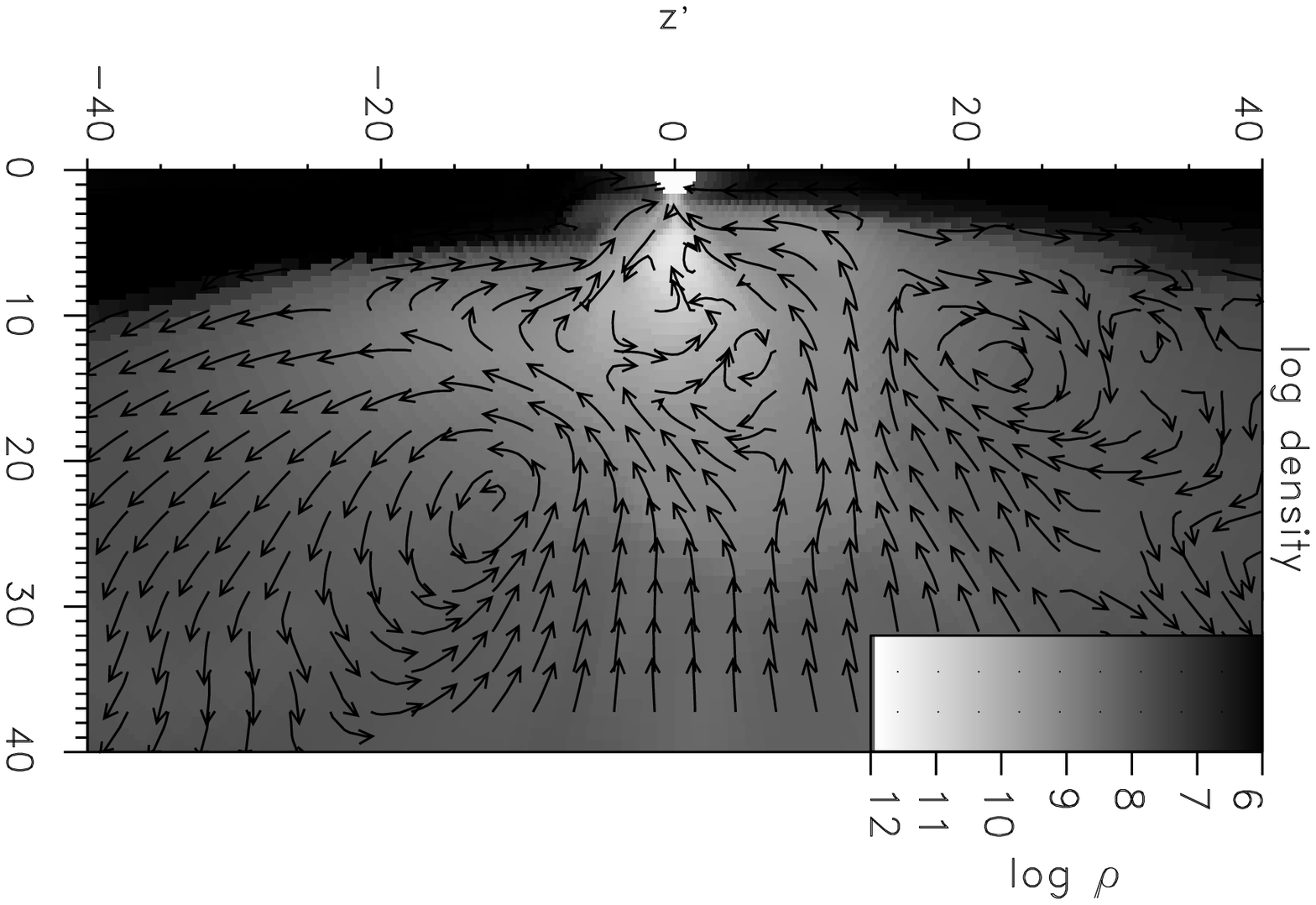}
  \includegraphics[bb=-50 590 500 100, width=.6\textwidth, angle=90]{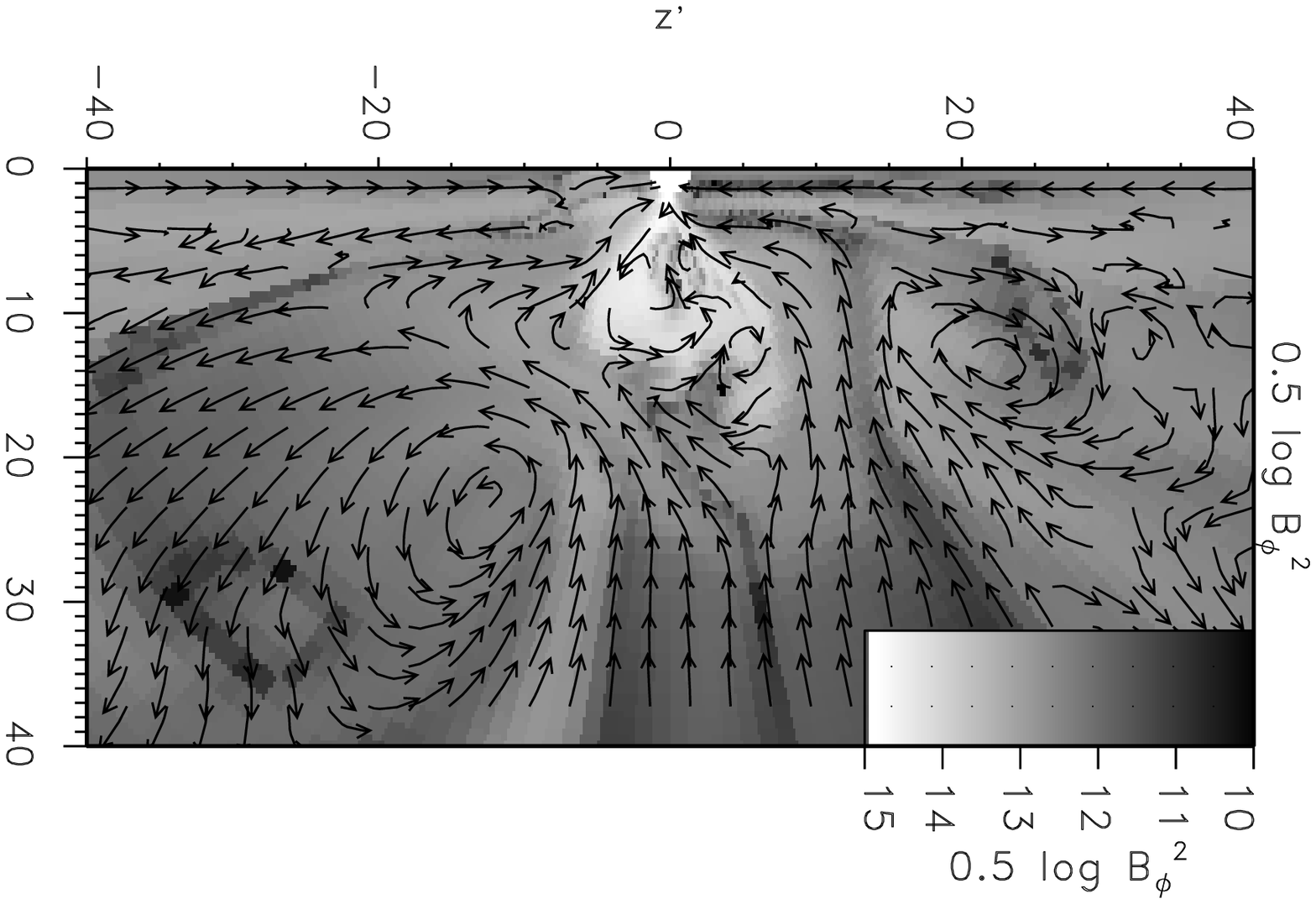}
  \includegraphics[bb=-50 910 500 100, width=.6\textwidth, angle=90]{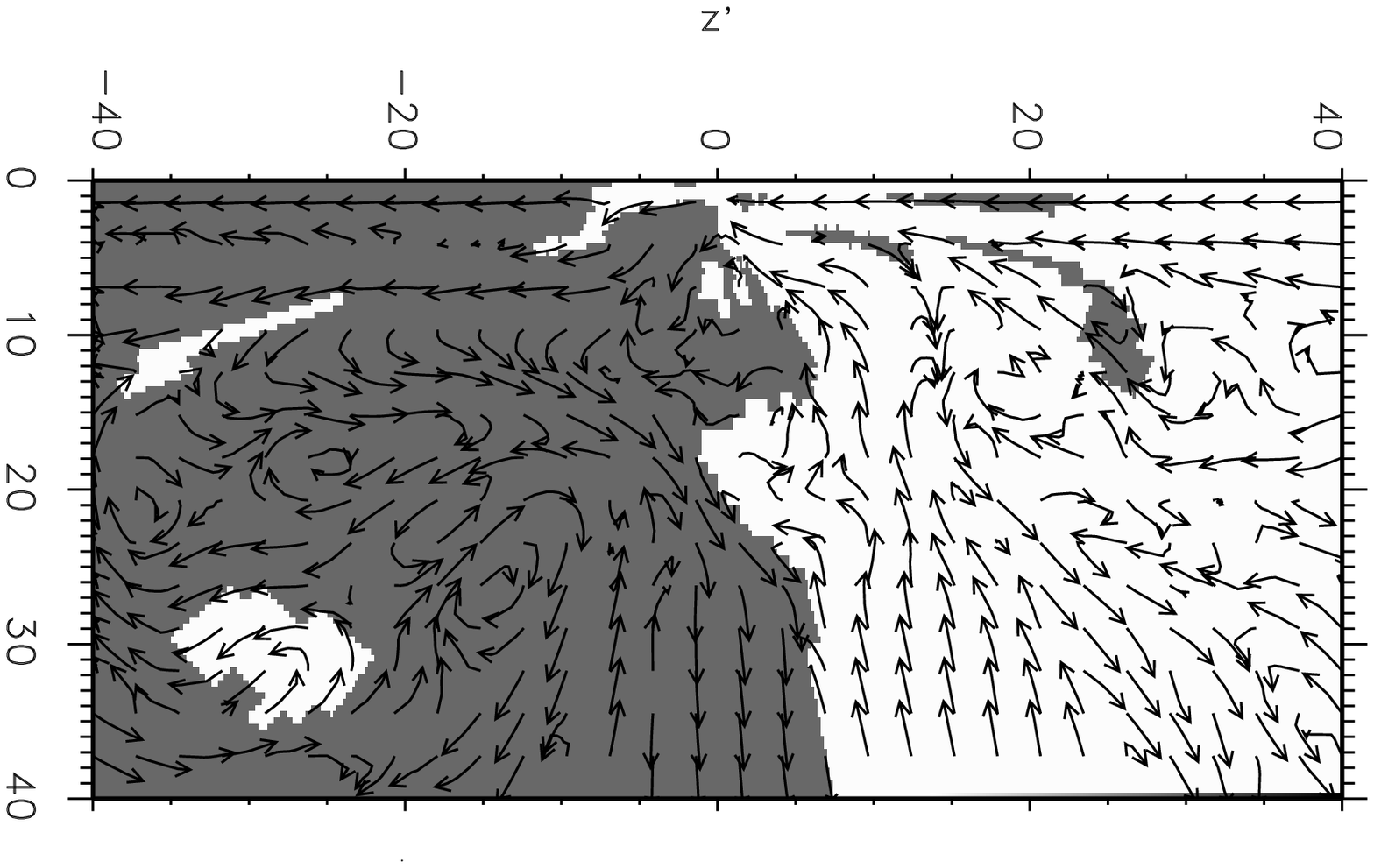}
  \caption{Maps of logarithmic density, toroidal magnetic field, 
and toroidal magnetic field domains with different
polarity (white and grey regions) for Proga et al.'s GRB simulation
at $t=0.2437$~s (left, middle, and right panel, respectively).
The arrows in the left and middle panels indicate
the direction of the poloidal velocity while the arrows in right panel
indicate the direction of the poloidal magnetic field.
The length scale is in units of the BH radius (i.e., $r'=r/R_S$
and $z'=z/R_S$}
\end{figure}

\begin{figure}
  \includegraphics[bb=-50 400 500 100, width=.6\textwidth, angle=90]{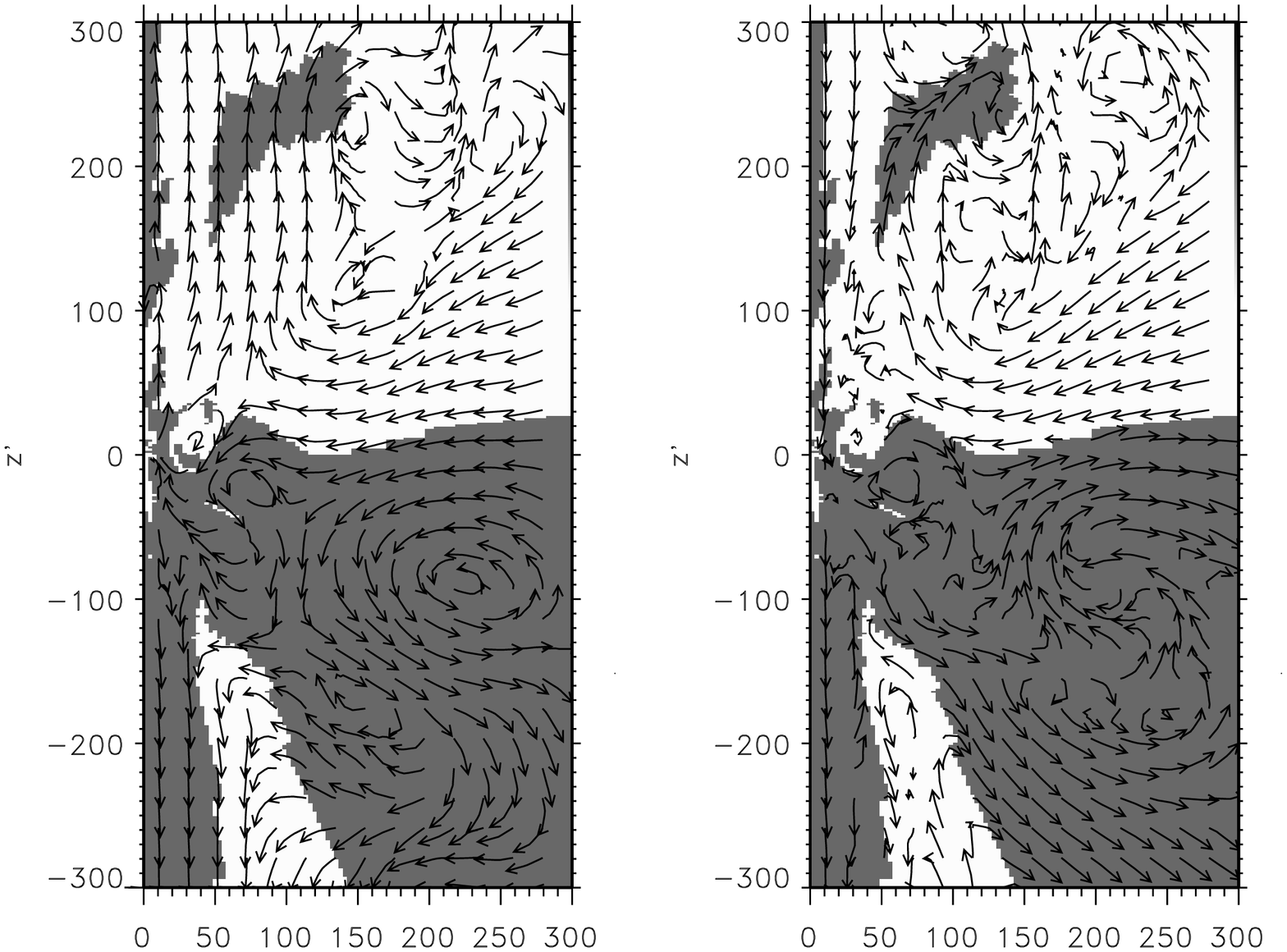}
  \caption{Maps of  toroidal magnetic field domains with different
polarity (white and grey regions) overplotted with the direction
of the poloidal velocity (left panel) and the direction
of the poloidal magnetic field (right panel) at the end of 
Proga \& Begelman's simulation of adiabatic accretion onto 
a BH (run D in \cite{PB03}).
The length scale is also in units of the BH radius; the same as in Fig. 1.
Note however the difference in the $r'$ and $z'$ ranges.}
\end{figure}

Figure~1 shows the flow pattern of the inner part of the flow at
$t=0.2437$~s. The left and middle panels show density and $|B_\phi|$ maps, 
respectively.  The two maps are overlaid with the direction of 
the poloidal velocity. The right panel shows the flow domains of 
different polarity of $B_\phi$ overlaid
with the direction of the poloidal magnetic field.
The polar regions of low
density and high $B_\phi$ coincide with the region of an outflow. We
note that during the latter phase of the evolution not all of the
material in the outflow originated in the innermost part of the torus
-- a  part of the outflow is ``peeled off'' the infalling
gas at large radii by the magnetic pressure (see Fig. 2 in 
\cite{Petal}).  
We imposed a weak poloidal field of a given polarity at the outer 
boundary. This means that {\it initially} $B_\phi$ changes sign only across
the equator, which is a relatively  unfavorable configuration for
subsequently producing $B_\phi$ reversals in the jet.  
Nevertheless, we find that the polarity of $B_\phi$ 
changes with time. This is because 
the flow loses memory of the initial polarity by the time it reaches the
inner MRI-dominated regions where the jet is formed. 

Sikora et al. (2003) \cite{Setal} argue that Poynting
flux-dominated jets with reversing B-fields provide a natural and efficient
way to dissipate energy via the reconnection process.
Poynting flux dominated jets have been found in previous numerical
simulations whereas jets with reversing B-fields appear as a relatively new result.
Therefore, we have reviewed results from \cite{PB03} to check whether
polar outflows generated during adiabatic accretion onto BHs also
exhibit reversing $B_\phi$. 
Simulations reported by \cite{PB03} are suitable to study the flow 
pattern on relatively large length scales because they 
have been continued for a few  dynamical time scales
at a distance of $\sim 10^3$ BH radii.

Fig. 2 shows the flow domains of different polarity of $B_\phi$
overlaid with the direction of the poloidal velocity and magnetic field
(left and right panel, respectively) for run D in \cite{PB03}.
It appears then that reversing B-fields are not unique to our GRB
simulations
and that domains with different polarity can be relatively large
and long-lived.

\section{Conclusions}
We have performed time-dependent two-dimensional MHD simulations of
the collapsar model.  Our simulations show that: 
1) soon after the
rotationally supported torus forms, the magnetic field very quickly
starts deviating from purely radial due to MRI and shear. This leads
to fast growth of the toroidal magnetic field as field lines wind up
due to the torus rotation; 
2) The toroidal field dominates over the
poloidal field and the gradient of the former drives a polar outflow
against supersonically accreting gas through the polar funnel; 
3) The polar outflow is Poynting flux-dominated; 
4) The polarity of the toroidal field can change with time;
5) The polar outflow reaches
the outer boundary of the computational domain ($5\times10^8$~cm) with
an expansion velocity of 0.2 c; 
6) The polar outflow is in a form of
a relatively narrow jet (when the jet breaks through the outer
boundary its half opening angle is $5^\circ$); 
7) Most of the energy
released during the accretion is in neutrinos, $L_\nu=2\times
10^{52}~{\rm erg~s^{-1}}$. 
Neutrino driving will increase the outflow energy (e.g., \cite{FM03}
and references therein), but could also increase
the mass loading of the outflow if the energy is deposited in the torus.


\begin{theacknowledgments}
DP acknowledges support from NASA under LTSA grants NAG5-11736 
and NAG5-12867.
MCB acknowledges support from NSF grants 
AST-9876887 and AST-0307502.
\end{theacknowledgments}


\bibliographystyle{aipproc}   

\bibliography{sample}

\IfFileExists{\jobname.bbl}{}
 {\typeout{}
  \typeout{******************************************}
  \typeout{** Please run "bibtex \jobname" to optain}
  \typeout{** the bibliography and then re-run LaTeX}
  \typeout{** twice to fix the references!}
  \typeout{******************************************}
  \typeout{}
 }

\end{document}